# Observation of Γ-valley moiré bands and emergent hexagonal lattice in twisted transition metal dichalcogenides

Ding Pei[1,2,9,*], Binbin Wang[1,7,*], Zishu Zhou[3,*], Zhihai He[4,8,*], Liheng An[3], Shanmei He[2], Cheng Chen[2], Yiwei Li[1], Liyang Wei[1], Aiji Liang[1], Jose Avila[5], Pavel Dudin[5], Viktor Kandyba[6], Alessio Giampietri[6], Mattia Cattelan[6], Alexei Barinov[6], Zhongkai Liu[1,9], Jianpeng Liu[1,9], Hongming Weng[8,7,4], Ning Wang[3,†], Jiamin Xue[1,†], Yulin Chen[2,1,9,†]

[1]*School of Physical Science and Technology, ShanghaiTech University, Shanghai 201210, China*
[2]*Department of Physics, University of Oxford, Oxford, OX1 3PU, United Kingdom*
[3]*Department of Physics, the Hong Kong University of Science and Technology, Clear Water Bay, Hong Kong, China*
[4]*Songshan Lake Materials Laboratory, Dongguan, Guangdong 523808, China*
[5]*Synchrotron SOLEIL, L'Orme des Merisiers, Saint Aubin-BP 48, 91192 Gif sur Yvette Cedex, France*
[6]*Elettra-Sincrotrone Trieste, Trieste, Basovizza 34149, Italy*
[7]*School of Physical Sciences, University of Chinese Academy of Sciences, Beijing 100190, China*
[8]*Beijing National Laboratory for Condensed Matter Physics, and Institute of Physics, Chinese Academy of Sciences, Beijing 100190, China*
[9]*ShanghaiTech Laboratory for Topological Physics, ShanghaiTech University, Shanghai 200031, China*

*These authors contributed equally to this work.
†Email: yulin.chen@physics.ox.ac.uk; xuejm@shanghaitech.edu.cn; phwang@ust.hk

**Twisted van der Waals heterostructures have recently been proposed as a condensed-matter platform for realizing controllable quantum models due to the low-energy moiré bands with specific charge distributions in moiré superlattices. Here, combining angle-resolved photoemission spectroscopy with sub-micron spatial resolution (μ-ARPES) and scanning tunneling microscopy (STM), we performed a systematic investigation on the electronic structure of 5.1° twisted bilayer $WSe_2$ that hosts correlated insulating and zero-resistance states. Interestingly, contrary to one's expectation, moiré bands were observed only at Γ-valley but not *K*-valley in μ-ARPES measurements; and correspondingly, our STM measurements clearly identified the real-space honeycomb- and Kagome-shaped charge distributions at the moiré length scale associated with the Γ-valley moiré bands. These results not only reveal the unusual valley dependent moiré-modified electronic structure in twisted transition metal dichalcogenides, but also highlight the Γ-valley moiré bands as a promising platform for exploring strongly correlated physics in emergent honeycomb and Kagome lattices at different energy scales.**

# I. INTRODUCTION

Moiré superlattices are formed when two layers of van der Waals (vdW) materials with a slight twist angle or lattice mismatch are stacked together. In such a superlattice, original electronic bands of the constituent layer are folded into a reduced Brillouin zone (BZ), or mini-BZ, due to the enlarged moiré lattice constant, which can give rise to isolated flat moiré bands under suitable conditions (e.g., specific twist angle or lattice relaxation, etc.) [1-5]. At proper filling level, these moiré bands can lead to rich interesting quantum phases, such as correlated insulator [6-10], superconductor [7,8], or quantized anomalous Hall insulator [11].

In real space, a complementary view of moiré bands is the electron localization on certain positions in the moiré superlattice, forming 'moiré orbitals' [5,12]. Depending on the geometry of moiré orbitals, a low energy moiré band can be effectively described by a paradigmatic quantum model at the moiré length scale. As correlation effects and topological properties of these moiré bands can be effectively tuned by various external parameters (e.g., strain [13], substrate [14], and displacement field [9]), vdW heterostructures with moiré superlattices can be regarded as a controllable 'quantum simulator' to realize different model Hamiltonians and explore their novel properties [12]. For example, it was proposed that the flat moiré bands in the 'magic-angle' twisted bilayer graphene have an equivalent real-space description as a honeycomb lattice at the moiré length scale [15-18], based on which intriguing correlated and topological states can be realized [6-8,11].

Apart from graphene-based moiré systems, twisted transition metal dichalcogenide (TMD) moiré superlattices are also gifted the versatility for studying rich emergent phases associated with different quantum systems (e.g., triangular, honeycomb and Kagome lattices) [19-22]. For examples, the $K$-valley moiré bands can be utilized to study triangular Hubbard model [20,21], while the $\Gamma$-valley moiré bands are related to the honeycomb and Kagome models [22]. Although the strongly correlated states in twisted TMDs have been revealed by recent transport results in twisted bilayer [9,10] and double-bilayer $WSe_2$ devices [23], a direct visualization of the moiré bands in the momentum space and the associated charge distribution in the real space is still lacking.

In this work, we performed a systematic investigation on the electronic structure of 5.1° twisted bilayer $WSe_2$ (tbWSe$_2$), which was reported to host moiré-induced zero-resistance states [9]. By using angle-resolved photoemission spectroscopy with sub-micron spatial resolution (μ-ARPES), electronic structures from both $\Gamma$- and $K$-valleys are directly observed. Interestingly, no sign of flat bands around $K$-valley was seen; instead, replica-like moiré bands around $\Gamma$-valley were clearly observed. Correspondingly, our scanning tunneling microscopy/spectroscopy (STM/STS) measurements revealed real-space charge distributions of the $\Gamma$-valley moiré bands at different energies, manifested as emergent honeycomb and Kagome lattices at the moiré length scale. The underlined physics of these observations can be well captured by our density functional theory (DFT) and tight-binding calculations, showing that moiré potential in the twisted TMD system can significantly modify the $\Gamma$-valley bands, making them a unique platform for simulating different quantum systems such as honeycomb and Kagome lattice models at different energy scales.

## II. RESULTS

The experimental layout and the device geometry used in this work are illustrated in Fig. 1(a). The tbWSe$_2$ sample was fabricated using standard tear-and-stack method (details can be found in the Methods section and Supplementary Information, SI), which was then transferred onto the substrate composed of layers of graphite, hBN and SiO$_2$/Si (Fig. 1(a)) for flatness and electric conductivity necessary for μ-ARPES and STM measurements. The optical image of the real device is shown in Fig. 1(b); and the 5.1° twist angle was confirmed by the STM measurement that shows a moiré periodicity of $\lambda \sim 3.7$ nm (Fig. 1(c)), as $\lambda = \frac{a}{2\sin(\theta/2)}$ where the lattice constant of WSe$_2$ was measured as $a = 0.33$ nm (see SI for more details).

A valley-resolved investigation on the electronic structure of tbWSe$_2$ was first performed with μ-ARPES. The consistency between the scanning photoemission microscopy (SPEM) map (Fig. 2(a)) and the optical image (Fig. 1(b)) enables us to accurately locate the measurement region on tbWSe$_2$. Contrary to the previous proposal that the strongly correlated states in tbWSe$_2$ could arise from the $K$-valley moiré flat band, no apparent sign of moiré band around the $K$-valley (of the top WSe$_2$ layer, labelled as $K_t$) is observed, neither in the equal energy contour mapping (Fig. 2(b)) nor in the energy-momentum dispersion across the $K_t$ point (Fig. 2(c)). The fact that no moiré flat bands were observed at $K$-valley from different measurements with various photon energies and sample orientations (see Fig. 2(d) and more in SI) rules out the suppression of the moiré bands caused by the matrix-element effect.

In contrast, additional features are observed both inside and outside the bare Γ-valley pocket (Fig. 2(b)) and dispersions (Figs. 2(e), (f)): replica-like moiré bands (Figs. 2(e), (f), marked as MBs) can be clearly decerned in addition to the intensive main Γ-valley bands (Figs. 2(e), (f), marked as Γ$_1$ and Γ$_2$) due to interlayer coupling of tbWSe$_2$. In Fig. 2(e), the momentum separation ($\delta k$) between the replica and main bands is determined to be $0.19 \pm 0.03$ Å$^{-1}$, which is in consistent with the reciprocal vector of 5.1° tbWSe$_2$ superlattice ($\sim 0.19$ Å$^{-1}$) but significantly smaller than that of the WSe$_2$/graphite heterostructure ($\sim 0.89$ Å$^{-1}$, see SI) confirming the tbWSe$_2$ origin of the replica band. Other possible origin of the replica bands can also be excluded, such as the domain effect (these replica bands have been consistently observed on different positions, see SI) or Umklapp scattering [24,25] of the bottom layer electronic states (as the band maximum of the monolayer WSe$_2$ Γ-valley is significantly lower than the MB [26]).

To uncover the fine structure of the moiré bands, next we carry out STS measurements on the same sample. The d$I$/d$V$ spectra from the constant-height scanning tunneling spectroscopy (CH-STS) along high-symmetry directions of the moiré lattice (AA-B-BR-B-AA, see the caption of Fig. 3(a) for definition) and three representative d$I$/d$V$ curves (on AA, B, BR sites) are illustrated in Fig. 3(b), which show multiple sharp d$I$/d$V$ peaks around the bias voltage $V_{bias}$ = -1.15 V on B and BR sites (Fig. 3(b)). As the d$I$/d$V$ signal decays exponentially with the increase of in-plane momentum of the tunneling electrons, such sharp peaks in CH-STS curves indicate the existence of flat moiré bands in the vicinity of the BZ center [27-30].

In order to extract the K-valley band information, we further carried out STS measurement in constant-current mode (CC-STS), which is more sensitive to tunneling electrons from $K$ valleys [27]. As expected, $dI/dV$ peaks absent in the CH-STS measurement around $V_{bias}$ = -1.0 eV are now visible as broad humps, showing the $K$-valley bare bands of WSe$_2$ layers (see SI for identification methods) but no sign of sharp peaks as those observed from the Γ-valley at $V_{bias}$= -1.15 V (visible in both CH-STS and CC-STS curves).

Interestingly, if we map out the local density of state (LDOS) distribution in real space at different $dI/dV$ peak energies (V$_1$-V$_4$, as marked in Fig. 3(b)), one can see different moiré orbital symmetry. As shown in Figs. 3d-g, the LDOS maps from STM measurements where bias voltages were set to $dI/dV$ peaks from AA (V$_4$, Fig. 3(g)) and B (V$_2$, Fig. 3(e)) sites both show honeycomb lattice, while the LDOS maps with $V_{bias}$ set to the $dI/dV$ peaks from BR site (V$_1$ for Fig. 3(d) and V$_3$ for Fig. 3(f)) clearly show Kagome type patterns. The observed charge distribution complies with a D$_6$ symmetry at the moiré length scale instead of the D$_3$ symmetry of the tbWSe$_2$ moiré superlattice as a result of the emergent D$_6$ symmetry of moiré potential, since the z to -z (the out-of-plane direction) reflection in lattice structure does not affect the overall moiré potential of a bilayer system [22]. (This is supported by the qualitatively identical $dI/dV$ spectra on the neighboring B sites with inversed configurations (Fig. 3(b))). According to the continuum theory [22], the emergence of Kagome lattice in twisted TMDs is related with the $sd2$ hybridization of moiré orbitals on Bernal stacking regions, which moves the Wannier centers from neighboring B sites to the middle point, i.e., BR sites, thus turning a honeycomb lattice to a Kagome lattice.

To understand above experimental observations, we performed both first-principles and tight-binding calculations. Notations of high symmetry points in the mini-BZ of tbWSe$_2$ superlattice are illustrated in Fig. 4(a). In the first-principles calculation (Fig. 4(b), left part), we find a set of less dispersive bands (highlighted in red) located around 0.25 eV below the valence band maximum (VBM). The calculated charge distributions corresponding to these bands appear as honeycomb or Kagome patterns at different binding energies (Fig. 4(d), right part; e$_1$ for the honeycomb pattern, e$_2$ for the Kagome pattern), which qualitatively agrees with our observations. To validate the origin of these flat bands, the unfolding procedure based on the atomistic tight-binding model has been utilized [31,32]. Unfolding the moiré band structures can be simply understood as the projection of calculated moiré wave functions onto wave functions of each monolayer WSe$_2$, which presents the weighted (based on the projection) moiré bands in the original BZ (as illustrated in Fig. 4(c)). In the tight-binding calculation (Fig. 4(d), left part), qualitatively equivalent moiré bands (highlighted in red) with similar range of binding energy are observed, confirming the consistency between the first-principles and tight-binding calculations. In the unfolded band structure along $K$-Γ-$M$ direction (Fig. 4(c), right part), we can immediately find these flat bands are the top-most Γ-valley moiré bands (the equivalent energies to e$_1$ and e$_2$ are indicated by the red dashed lines). Thus, we conclude the honeycomb and Kagome models are intrinsic properties of Γ-valley moiré bands.

## III. DISCUSSION & CONCLUSTION

While the absence of $K$-valley flat bands in both μ-ARPES and STM/STS measurements (the Γ-valley moiré bands are clearly observed) may seem unexpected, it could be understood from the $K$-valley orbital components [33]: in each constituent WSe$_2$ layer, the $K$-valley bands are contributed by orbitals with in-plane orientation (metal's $d_{xy}$, $d_{x^2-y^2}$ and chalcogen's $p_x$, $p_y$ orbitals), which are weakly coupled between layers due to the small interlayer overlap of their wavefunctions; while for Γ-valley, the bands are formed by metal's $d_{z^2}$ and chalcogen's $p_z$ orbitals, which naturally have larger overlap (thus interlayer coupling). Indeed, the continuum theory predicts the moiré potential in $K$-valley model is less than one-fifth of that in Γ-valley model [20-22]. Displacement field, which is indispensable in realizing strongly correlated phases in twisted TMDs, could be the key factor for the enhancement of moiré potential effects on the $K$-valley bands. Further ARPES and STM experiments with electrostatic gating are needed to search for the $K$-valley moiré bands in 5.1° tbWSe$_2$.

The Γ-valley moiré bands have narrow band widths (see Figs. 3(b), (c)) favorable for hosting strongly correlated phases. Moreover, the fact that the real space LDOS mapping at the d$I$/d$V$ peak energy shows different (honeycomb- and Kagome-shaped) charge distribution at the moiré length scale indicates the Γ-valley moiré bands can serve as a flexible platform to realize controllable honeycomb and Kagome quantum model systems, which may host rich correlated and topological states at proper fillings [34,35]. This scenario is experimentally feasible, as the majority of 2H-phase TMD homo-multilayers (e.g., WS$_2$, MoS$_2$ and MoSe$_2$) [26,36,37] and hetero-multilayers (e.g., MoS$_2$/WS$_2$ bilayer) [38] have their VBM at the Γ point, and thus the Γ-valley moiré bands would dominate their electric properties via electrostatic gating. (The Γ-valley moiré bands of tbWSe$_2$ can also be pushed up to the VBM when in-plane lattice constant increases by ~2.5%, see SI.) Indeed, recent transport measurements on twisted double-bilayer WSe$_2$ whose VBM is at the Γ point [23], have already revealed strongly correlated phases result from the Γ-valley moiré bands.

In conclusion, our results point out the importance of the Γ-valley moiré bands in the twisted TMDs, which gives a new direction for the exploration of correlated states in honeycomb and Kagome models.

## ACKNOWLEDGEMENTS

B. W. and J. X. acknowledge financial support from the Ministry of Science and Technology of China (2017YFA0305400) and the Strategic Priority Research Program of Chinese Academy of Sciences (XDA18010000), and experimental support from the Soft Nano Fabrication Center at ShanghaiTech. D. P. and S. M. H. thank the support from China Scholarship Council. Y. W. L. acknowledges the support from the National Natural Science Foundation of China (No. 12104304) and China Postdoctoral Science Foundation (No. 2021M692131). H. M. W. acknowledges the support from the National Natural Science Foundation of China (Nos. 2018YFA0305700, XDB33000000, 11925408, 12188101). Z. S. Z.,

L. H. A. and N. W. thank the support from the National Key R&D Program of China (2020YFA0309600) and the Hong Kong Research Grants Council (Project Nos. 16303720). D. P. thanks the useful discussion with N. B. M. S. The micro-ARPES research in SOLEIL is under the proposal 20210047. The research leading to micro-ARPES results in Elettra under the proposal 20195301, has been supported by the project CALIPSO+ under Grant Agreement 730872 from the EU Framework Program for Research and Innovation HORIZON 2020. Preliminary ARPES experiments were performed at nano-ARPES branch of BL07U endstation, SSRF. The computational resource is provided by the Platform for Data-Driven Computational Materials Discovery in Songshan Lake material Laboratory.

## APPENDIX: METHODS

### 1. Sample preparation

The tbWSe$_2$ device was fabricated by using dry-transfer technique. The monolayer and few layers WSe$_2$ flakes were mechanically exfoliated from bulk crystal onto the SiO$_2$ wafer. Polycarbonate film was used to tear part of the target monolayer WSe$_2$ at around 80 °C to 100 °C. The left part of WSe$_2$ was rotated manually by a twist angle ($\theta$) around 5.1° ($\pm$ 0.1°) and stacked together. For the substrate part, few layers boron nitride films with a thickness of 10 nm to 20 nm were exfoliated on the silicon wafer. Then 4-5 layers graphite thin films were picked up by polycarbonate film and transferred onto the prepared boron nitride. Then the tbWSe$_2$ was transferred onto the graphite thin film. Finally, a Cr/Au electrode of 5 nm/7 nm was deposited on part of the graphite films to form the contact.

### 2. μ-ARPES measurement

Samples were annealed at 300 ℃ for 4 hours before ARPES measurements. All experiments were performed with a base vacuum better than 3×10$^{-10}$ mbar. Experiments at the SpectroMicroscopy beamline of Elettra Sincrotrone Trieste were performed at 94 K with photon energies of 27 and 74 eV, polarization of linear horizontal (LH). The estimated energy and angular resolutions were ~ 50 meV and 0.3°, respectively. Experiments at the analysis nano-spot angle-resolved photoemission spectroscopy (ANTARES) beamline of Synchrotron SOLEIL, were performed at 75 K with a photon energy of 100 eV, polarization of LH. The estimated energy and angular resolutions were ~ 40 meV and 0.5°, respectively.

### 3. STM measurement

The STM and STS measurements were performed under ultrahigh vacuum (pressure $\leqslant$ 10$^{-11}$ mbar) and liquid-helium temperature with an Omicron low-temperature STM. The etched tungsten wires were used as the STM tip. The constant-height d$I$/d$V$ spectroscopy was acquired by turning off the feedback loop and using the standard lock-in techniques ($f$ = 991.7 Hz, $V_{Ampl}$ = 4 mV). The constant-current d$I$/d$V$ spectroscopy was acquired by leaving on the feedback loop. The tip-sample distance Z changed to keep the constant tunneling current. The tip was calibrated on Ag (111) before measurements on the samples. STM/STS data were analyzed using SPIP 6.7.3 and MATLAB.

# 4. Calculation

The first-principle calculations are performed by using the Vienna ab initio simulation package [39] with the projector-augmented wave potential method [40-42]. The exchange-correlation potential is described using the generalized gradient approximation in the Perdew-Burke-Ernzerhof form [43]. The energy cutoff of the plane-wave basis set is 300 eV. A vacuum region larger than 15 Å is applied to ensure no interaction between the slab and its image. In our optimization, all structures are relaxed until the force on each atom is less than 0.01 eV/Å. The Van der Waals interactions between the adjacent layers are considered by using zero damping DFT-D3 method of Grimme [44].

We unfold the bands of twisted bilayer to the primitive-cell BZ of a monolayer [31,32]. The tight-binding method is adopted to calculate the unfolded bands of twisted bilayer $WSe_2$. The band structure of monolayer $WSe_2$ can be described by a tight-binding Hamiltonian consisting of 11 atomic orbitals, the *d* orbitals for W and the *p* orbitals for Se. Here, we consider the first-neighbor W-W, Se-Se, W-Se hopping terms, and the second-neighbor W-Se hopping term to improve the accuracy. The tight-binding parameters are obtained by fitting the low energy conduction and valence bands, which are given in Table S1. The effect of spin-orbit coupling (SOC) is included by adding an on-site term $\sum_\alpha \lambda_\alpha \boldsymbol{L} \cdot \boldsymbol{S}$ in the Hamiltonian, where $\alpha$ stands W or Se atom. The parameters of SOC terms are taken from ref. [45]. More details are available in SI.

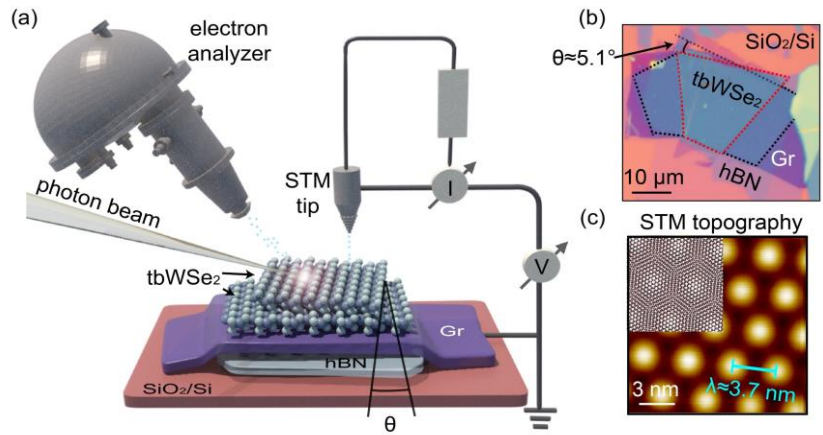

FIG. 1. Device design and experiment layout. (a) Schematic illustration of the device design and the experiment layout. From bottom to top, the red, light-blue, purple sheets stand for $SiO_2/Si$, hBN and graphite substrates, respectively; the atomic model stands for a bilayer $WSe_2$ with a twist angle ($\theta$); for ARPES measurements, photoelectrons are collected by a hemisphere analyzer; for STM measurements, tunnelling electrons are tuned by the current ($I$) and voltage ($V$) sources and collected by the STM tip. (b) Optical image of 5.1° $tbWSe_2$ device. Dashed lines of red and black mark boundaries of top and bottom layer $WSe_2$, respectively. $SiO_2/Si$, hBN and graphite substrates appear in the same colors as shown in the illustration. The included angle of upper boundaries of top and bottom layers is the twist angle, $\theta \approx 5.1°$. (c) STM topography, acquired at fixed bias voltage $V_{bias} = -1.39$ V and current $I = 90$ pA, showing the moiré pattern with a periodicity of $\lambda \approx 3.7$ nm. The inset illustration presents the atomic structure of $tbWSe_2$.

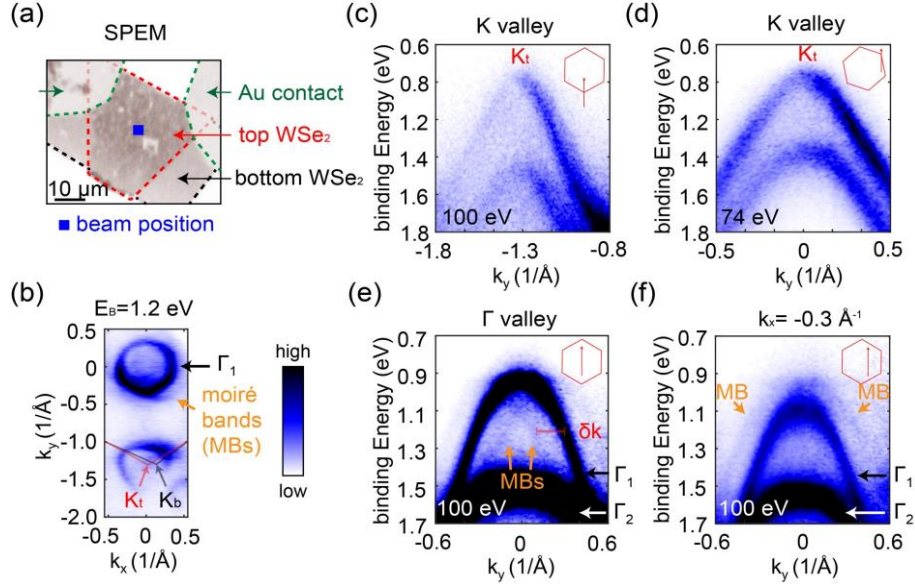

FIG. 2. ARPES measurement on tbWSe$_2$. (a) Scanning photoemission microscopy (SPEM) of tbWSe$_2$ device. The SPEM image is generated from real space mapping of the valence band spectra. Dashed lines of red, black, and green mark boundaries of top, bottom layer WSe$_2$, and Au contact, respectively. The blue square marks the photon beam position during ARPES measurements. (b) Equal energy contour at binding energy $E_B$=1.2 eV measured at 100 eV, overlapped with BZs of the top (red) and bottom (black) layer WSe$_2$. The subscripts 't' and 'b' of $K$-valleys stand for the top and bottom layer, respectively. The bare Γ-valley pocket (of main band $Γ_1$) is indicated by the black arrow. Moiré bands (MBs) are indicated by the orange arrow. (c) Band dispersion across $K_t$ point measured at 100 eV. (d) Band dispersion across $K_t$ point measured at 74 eV. (e) Band dispersion across Γ points measured at 100 eV. (f) Band dispersion along $k_y$ at $k_x$= -0.3 Å$^{-1}$ (around Γ-valley) measured at 100 eV. In (c)-(f), the inset red hexagons are the BZs of top layer WSe$_2$, showing the sample orientation; red arrows present the directions of dispersions with respect to BZs. In (e) and (f), $Γ_1$ and $Γ_2$ stand for two main bands of bilayer WSe$_2$; MBs are indicated by orange arrows; $δk$ is the momentum spacing between MBs and $Γ_1$.

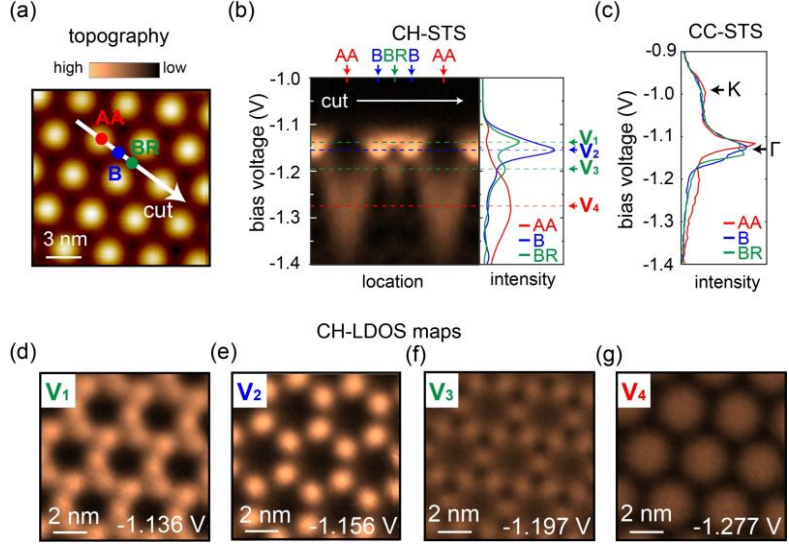

FIG. 3. STM measurement on tbWSe$_2$. (a) Definition of high symmetry sites in tbWSe$_2$ superlattice. Here, AA (indicated by the red point) means eclipsed stacking sites with W atoms over W atoms; BR (indicated by the green point) means the bridge that connects neighboring AA sites; B (indicated by the blue point) stands for Bernal stacking, including two inversed configurations with W atom over Se atom (B$_{W/Se}$) or Se atom over W atom (B$_{Se/W}$). The white arrow indicates the direction of d$I$/d$V$ spectra shown in (b). (b) Left: d$I$/d$V$ spectra measured in constant height mode (CH-STS) along AA-B-BR-B-AA direction. The positions of AA, B, BR sites are indicated by arrows of corresponding colors. Right: representative d$I$/d$V$ curves on AA (red), B (blue) and BR (green) sites. The peak positions in AA (V$_4$), B (V$_2$) and BR (V$_1$, V$_3$) curves are marked by dashed lines of corresponding colors. (c) d$I$/d$V$ curves measured in constant current mode (CC-STS) on AA, B and BR sites. Black arrows indicate d$I$/d$V$ signal contributed by $\Gamma$- and $K$-valleys, respectively. (d)-(g) LDOS maps (in CH mode) corresponding to the peak positions in the AA (V$_4$), B (V$_2$) and BR (V$_1$, V$_3$) curves, respectively. During the LDOS mapping, the tip height was adjusted at $V_{bias}=$ -1.6 V with $I=$ 170 pA.

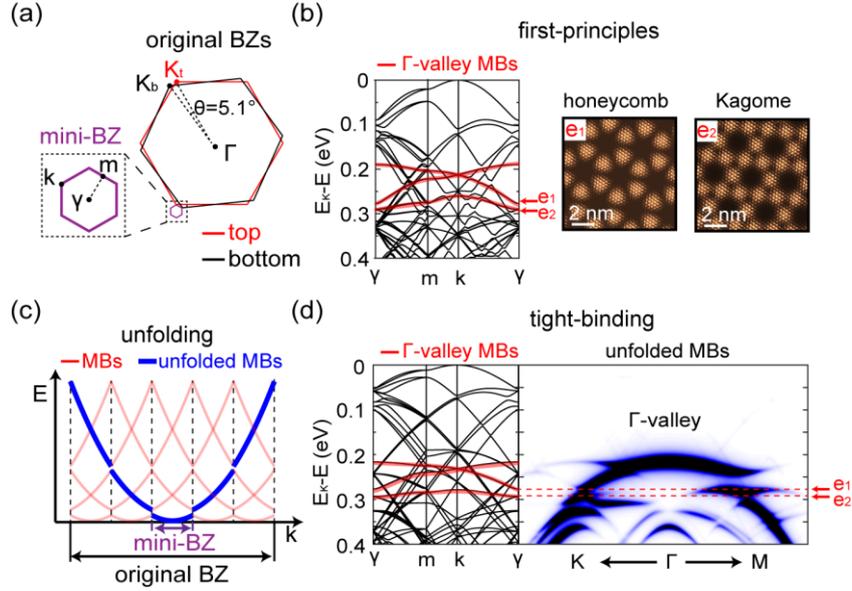

FIG. 4. Calculation on tbWSe$_2$. (a) Original BZs of a bilayer WSe$_2$ with a twist angle $\theta = 5.1°$ and resulting mini-BZ of tbWSe$_2$ superlattices. The BZs of top layer, bottom layer WSe$_2$ and the mini-BZ of tbWSe$_2$ are illustrated in red, black, and purple, respectively. High symmetry points are labelled on BZs; the subscripts 't' and 'b' stand for the top and bottom layer, respectively. (b) Left part: first-principles-calculated moiré bands (MBs) of 5.1° tbWSe$_2$ presented in the mini-BZ. The top-three Γ-valley MBs are highlighted in red. Right part: calculated charge distributions at energies e$_1$ and e$_2$ (indicated by red arrows in the left part), showing honeycomb (e$_1$) and Kagome patterns (e$_2$). (c) A simplified example illustrating the band unfolding in 1D (more discussion of 2D case can be found in SI). Blue and red lines show the unfolded and folded bands. (d) Left part: tight-binding-calculated MBs of 5.1° tbWSe$_2$ presented in the mini-BZ. The top-three Γ-valley MBs are highlighted in red. Right part: the unfolded MBs around Γ-valley (along $K$-Γ-$M$ direction in the original BZ). The equivalent positions of e$_1$ and e$_2$ are indicated by red arrows and dashed lines.

# Supplementary information for

# Observation of Γ-valley moiré bands and emergent hexagonal lattice in twisted transition metal dichalcogenides

Ding Pei[1,2,9,*], Binbin Wang[1,7,*], Zishu Zhou[3,*], Zhihai He[4,8,*], Liheng An[3], Shanmei He[2], Cheng Chen[2], Yiwei Li[1], Liyang Wei[1], Aiji Liang[1], Jose Avila[5], Pavel Dudin[5], Viktor Kandyba[6], Alessio Giampietri[6], Mattia Cattelan[6], Alexei Barinov[6], Zhongkai Liu[1,9], Jianpeng Liu[1,9], Hongming Weng[8,7,4], Ning Wang[3,†], Jiamin Xue[1,†], Yulin Chen[2,1,9,†]

[1]*School of Physical Science and Technology, ShanghaiTech University, Shanghai 201210, China*
[2]*Department of Physics, University of Oxford, Oxford, OX1 3PU, United Kingdom*
[3]*Department of Physics, the Hong Kong University of Science and Technology, Clear Water Bay, Hong Kong, China*
[4]*Songshan Lake Materials Laboratory, Dongguan, Guangdong 523808, China*
[5]*Synchrotron SOLEIL, L'Orme des Merisiers, Saint Aubin-BP 48, 91192 Gif sur Yvette Cedex, France*
[6]*Elettra-Sincrotrone Trieste, Trieste, Basovizza 34149, Italy*
[7]*School of Physical Sciences, University of Chinese Academy of Sciences, Beijing 100190, China*
[8]*Beijing National Laboratory for Condensed Matter Physics, and Institute of Physics, Chinese Academy of Sciences, Beijing 100190, China*
[9]*ShanghaiTech Laboratory for Topological Physics, ShanghaiTech University, Shanghai 200031, China*

*These authors contributed equally to this work.
†Email: yulin.chen@physics.ox.ac.uk; xuejm@shanghaitech.edu.cn; phwang@ust.hk

## 1. Sample fabrication

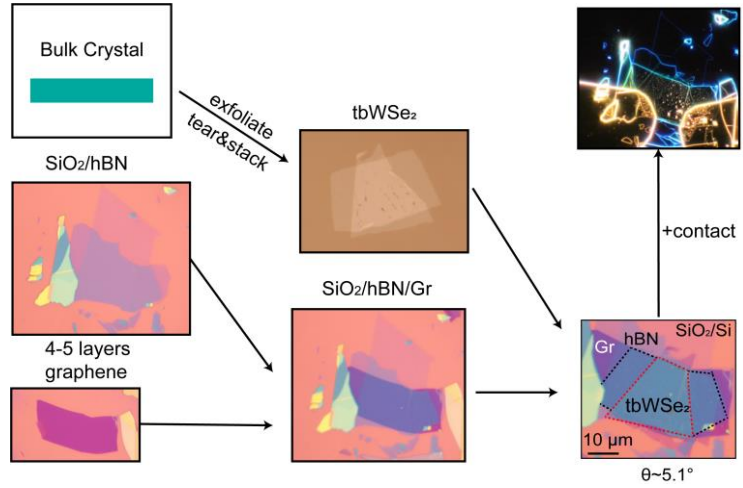

**Fig. S1| Sample fabrication procedure.**

## 2. Determination of the twist angle

The twist angle of the tbWSe$_2$ device was checked by both STM and ARPES measurements. From the fast Fourier transform (FFT) of the STM topography (Figs. S2 a,b), lattice constants of the WSe$_2$ monolayer and the tbWSe$_2$ superlattice are determined as $a = 0.33 \pm 0.02$ nm and $\lambda = 3.7 \pm 0.1$ nm, respectively. Using the relationship $\lambda = \frac{a}{2\sin(\theta/2)}$, we can obtain the twist angle as $5.1° \pm 0.1°$. The twist angle in ARPES data is determined by overlapping the Brillouin zones (BZs) of top (red) and bottom layers (black) WSe$_2$ onto the equal energy contour (Figs. S2 c,d). Here, we choose the $K$-valleys in the second BZ due to large twist-induced separation in the momentum space. Apart from the intensive $K$-valley of the top layer, the weak feature on the right side can be assigned to the $K$-valley of the bottom layer. The overlapping result reveals the twist angle as $5.1° \pm 0.2°$.

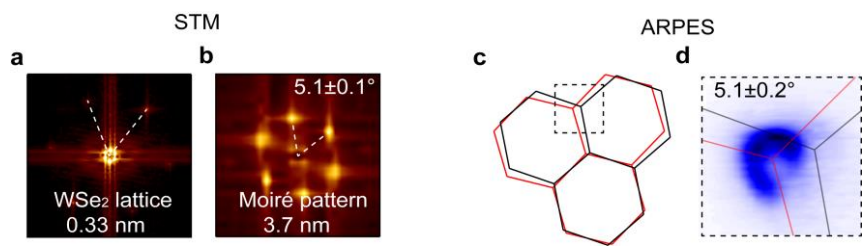

**Fig. S2| Determination of the twist angle. a**, Fast Fourier transform (FFT) of the STM topography. **b**, Zoom-in view of the central area in **a**. **c**, BZs of the top (red) and bottom (black) layer WSe$_2$. The location of $K$ valleys in the second BZ is highlight by the dashed square. **d**, ARPES equal energy contour around $K$ valleys in the second BZ, overlapped with BZs.

## 3. Additional ARPES data on the $K$-valley

As shown in Fig. S3, $K$ valley of the tbWSe$_2$ sample was measured with different in-plane orientation and photon energies (27 eV, 74 eV and 100 eV), though no obvious sign of flat band was found within the resolution of equipment.

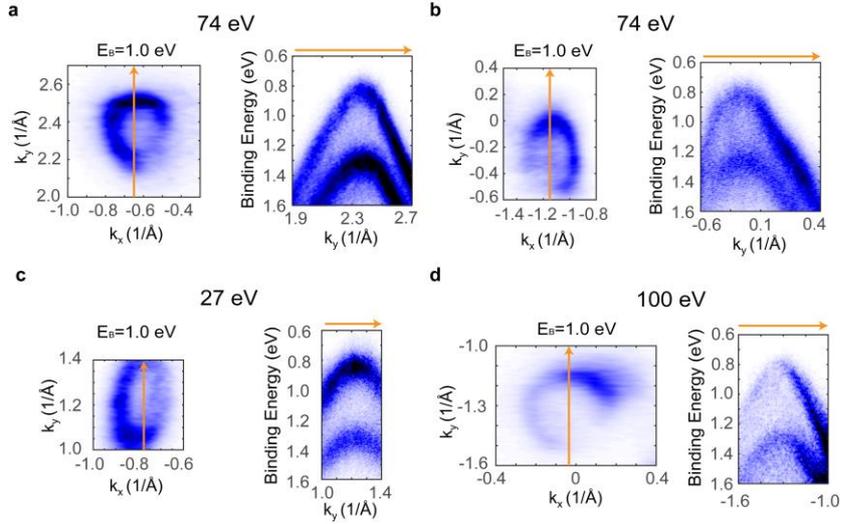

**Fig. S3| Additional ARPES data on the *K*-valley measured at different photon energies.** In **a-b** the left part shows the equal energy contour around the *K*-valley at $E_B$=1.0 eV; the right part shows the energy-momentum distribution along $k_y$ direction (indicated by the orange arrow). Data in **a-b** was obtained at 74 eV; data in **c** was obtained at 27 eV; data in **d** was obtained at 100 eV.

## 4. Additional ARPES data on the Γ-valley

Γ-valley moiré bands have been consistently observed on different positions in the 5.1° tbWSe$_2$ sample. Figure S4 (a) presents the same SPEM image as shown in the main text (Fig. 2a), where P1 (blue) marks the beam position for the data used in the main text (Figs. 2b-2f). The band dispersion along Γ-*K* direction acquired at P2 (green) presents moiré band both inside and outside the main bands (Fig. S4 (b)), consistent with the data acquired at P1 (Fig. 2e). The equal energy contour (Fig. S4 (c)) and band dispersion (Fig. S4 (d)) acquired at P3 (red) with 74 eV photons also show distinguishable signatures of Γ-valley moiré bands.

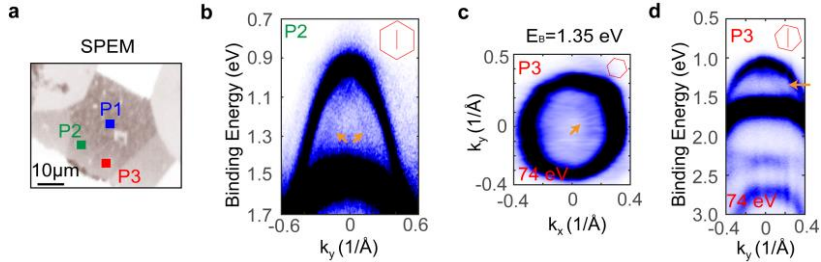

**Fig. S4| Moiré bands observed at different sample positions. a**, SPEM image showing the beam positions P1 (blue), P2 (green) and P3 (red). **b**, Band dispersion along Γ-K direction, acquired at P2. **c**, Equal energy contour at $E_B$=1.35 eV, acquired at P3. **d**, Band dispersion across Γ point, acquired at P3. In **b-d,** the inset red hexagons are the BZs of top layer WSe$_2$, showing the sample orientation; red arrows present the directions of dispersions with respect to BZs; MBs are indicated by orange arrows. Data in **b** was obtained at 100 eV; data in **c-d** was obtained at 74 eV.

The equal energy contour around Γ valley is presented in Fig. S5 a. In addition to the intensive main band pocket, the inside ring-shaped features and the outside petal-like features are clearly presented. The momentum distribution curve (MDC) along the $k_y$ direction at $k_x$=0 shows two prominent peaks of the main bands (indicated by the blue arrows) and multiple peaks of these additional features (indicated by the red arrows). Within the resolution of equipment, the momentum spacing ($\delta k$) between the main band and the additional band turns out to the same, which is 0.19±0.03 Å$^{-1}$. The moiré vector $G_{moiré}$ of 5.1° tbWSe$_2$ has a length of 0.196 Å$^{-1}$, consistent with $\delta k$. Besides, according to the geometric relation between original BZ and mini-BZs (Fig. S5 b), one of the directions of $G_{moiré}$ is very close to $k_y$. Based on these facts, we propose a geometric understanding for these observed additional features. (This proposal is experiment-derived, and its validity waits for further theoretical examination.) As illustrated in Fig. S5 c, the blue circle stands for the main band, the gray circle stands for the shifted main bands (replica bans) by a $G_{moiré}$. An inside ring (highlighted in red) is formed by the inner parts of replica bands shifted in all-six directions, and petal-like features are formed by the outside edges of the replica bands. The symmetric replica-like bands are immediately observed in the energy-momentum dispersions along the $k_y$ direction (Fig. S5 d). In the cuts away from the BZ center (Figs. S5 d(i) and (ii)), dispersions outside the main bands are clearly observed, which form the petal-like features. In the cuts close the BZ center (Figs. S5 d(iii)-(v)), additional bands both inside and outside the main bands are observed, which form the ring-shaped and petal-like features.

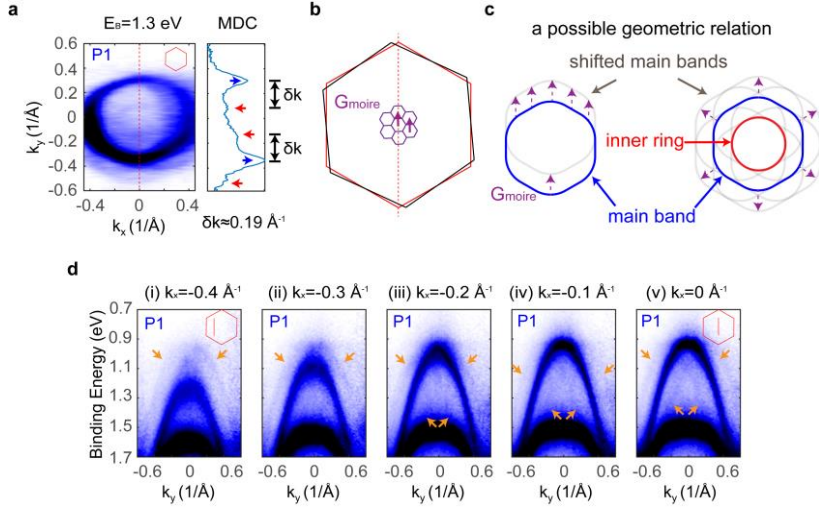

**Fig. S5| Additional ARPES data on the Γ-valley. a**, Equal energy contour at $E_B$= 1.3 eV and MDC at $k_x$=0 (indicated by the red dashed line), measured at P1. Red and blue arrows indicate the peak positions of moiré bands and main bands, respectively. $\delta$k marks the momentum separation between moiré bands and main bands, which is around 0.19 Å$^{-1}$. **b**, Geometric relation between original BZ, mini-BZs and k$_y$ direction. The red hexagon has the same orientation as that in **a**. **c**, Illustration of the first order replica bands (gray), generated by shifting the Γ-valley pocket via the moiré vector of tbWSe$_2$, $G_{moiré}$ (indicated by the purple arrow). Main bands, shifted main bands (replica bands), and the resulting inner ring are highlighted in blue, gray, and red, respectively. **d**, Band dispersions along $k_y$ (Γ-K direction) measured at (i)-(v) from $k_x$=-0.4 Å$^{-1}$ to $k_x$=0, respectively. In **a** and **c**, the inset red hexagons show the orientation of BZs. In **c**, red arrows present the directions of dispersions with respect to BZs; MBs are indicated by orange arrows.

## 5. Discussion on the substrate effect in ARPES

The twist angle between graphite substrate and bottom layer WSe$_2$ was determined as ~11° (Fig. S6 a), by overlapping BZs of WSe$_2$ (black for bottom layer, red for top layer) and graphite (green) onto the equal energy contour. The resulting BZ of WSe$_2$/Gr heterostructure (Fig. S6 b, blue) has a reciprocal vector length of ~0.89 Å$^{-1}$, significantly larger than the BZs of tbWSe$_2$ superlattice (Fig. S6 b, purple).

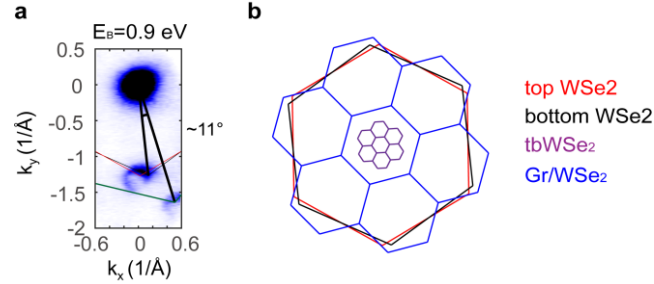

**Fig. S6| Discussion on substrate effect. a**, Equal energy contour at $E_B$=0.9 eV, overlapped with BZs of top (red) and bottom (bottom) layer WSe$_2$ and graphite (green). The twist angle between bottom layer WSe$_2$ and graphite substrate is determined to be 11°. **b**, BZs of top layer (red), bottom layer (black) WSe$_2$, mini-BZs of 5.1° tbWSe$_2$ (purple) and 11° Gr/WSe$_2$ superlattices (blue), respectively.

There are four routes via which graphite substrate could contribute to the ARPES signal around Γ-valley, namely: 1. The σ band of graphite (Γ pocket of graphite); 2. The replica graphite π bands (K pocket of graphite) induced by Gr/WSe$_2$ hybridization. 3. The replica WSe$_2$ Γ-valley bands induced by Gr/WSe$_2$ hybridization. 4. The replica WSe$_2$ K-valley bands induced by Gr/WSe$_2$ hybridization. Here, we discuss all possible situations case by case.

1. The σ band of graphite.

Previous ARPES measurements (figure S7 a, adapted from [1]) have revealed that the σ band of graphite is located 4 eV below the top of π band, which is far below the binding energy range (0.8 – 1.7 eV) discussed in this manuscript. Thus, we conclude the observed Γ-valley replica bands cannot originate from the σ band of graphite.

2. Replica graphite π bands.

Umklapp scattering in graphene/TMD heterostructure can result in replica bands of both graphite and TMD [2,3]. Considering the hybridization of all three constituents (figure S7 b), the original graphite π bands (illustrated as green circle) can be scattered via a reciprocal vector of bottom or top layer WSe$_2$ (black or red arrows), resulting in replica π bands inside the BZ of WSe$_2$ (illustrated as black and red circles). However, these graphite replica bands are far from the BZ center and can be clearly distinguished from the Γ-valley TMD bands, as confirmed by ARPES measurements (figure S7 c). Moreover, the shape of graphite π band is significantly different from that of TMD Γ-valley band. Thus, we conclude the observed Γ-valley replica bands cannot originate from the σ band of graphite.

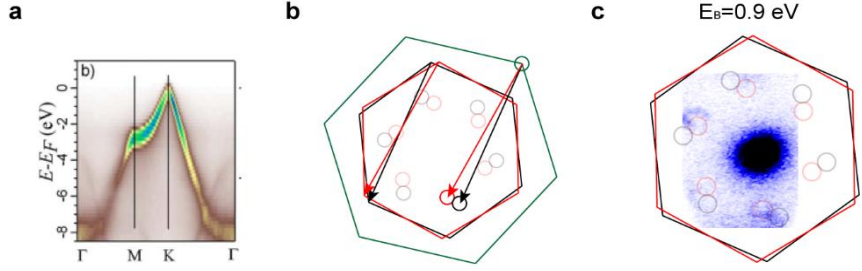

**Fig. S7| Graphite σ and π bands. a**, Band dispersion of graphite, adapted from [1]. **b**, BZs of top layer $WSe_2$, bottom layer $WSe_2$ and graphite are marked in red, black, green, respectively. Black and red arrows stand for reciprocal vectors of top and bottom layer $WSe_2$, respectively. The green circle stands for the K pocket of graphite, red and black circles are replica graphite K pockets. **c**, Equal energy contour at $E_B$=0.9 eV, overlapped with part of **b**, showing the positions of replica bands.

3. Replica $WSe_2$ Γ-valley bands induced by Gr/$WSe_2$ hybridization.

Significant difference in BZ sizes indicates that the hybridization between bottom layer $WSe_2$ and graphite substrate cannot result in the observed Γ-valley replica bands. Here, we explain this in detail.

Considering the hybridization of graphite and bottom layer $WSe_2$ (figure S8), the $WSe_2$ Γ-valley pocket in the second BZ (marked by black circle) can be scattered via a reciprocal vector of graphite (marked by green arrow). The resulted replica Γ-valley bands will be located in the center of mini-BZs of Gr/$WSe_2$ superlattice, the positions of which are far from the BZ center (figure S8). Therefore, the observed Γ-valley replica bands cannot result from the hybridization of graphite and bottom layer $WSe_2$. Using the same geometrical method, we can exclude the contribution by the hybridization between graphite and top layer $WSe_2$.

The hybridization between tb$WSe_2$ and graphite, if strong, will lead to at least two sets of replica bands (intrinsic tb$WSe_2$ replica moire bands and their replica induced by graphite), which is inconsistent with our observation.

Thus, we can conclude the observed Γ-valley replica bands cannot originate from the hybridization of graphite and $WSe_2$.

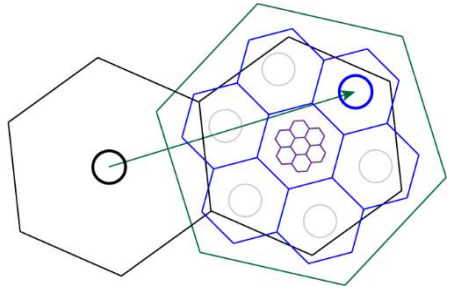

**Fig. S8| Γ-valley replica bands induced by Gr/$WSe_2$ superlattices.** BZs of graphite (green), bottom layer (black) $WSe_2$, mini-BZs of 5.1° tb$WSe_2$ (purple) and 11° Gr/$WSe_2$ superlattices (blue), respectively. The green arrow stands for the reciprocal vector of graphite. The black and blue circles stand for the original and replica Γ-valley bands, respectively.

4. Replica $WSe_2$ K-valley bands induced by Gr/$WSe_2$ hybridization.

Using the same geometrical method (figure S9), we can find the replica WSe$_2$ K-valley bands should also be located far from BZ center. Moreover, the shape of K-valley is significantly different from that of Γ-valley. Thus, we conclude the observed Γ-valley replica bands cannot originate from the WSe$_2$ K-valley.

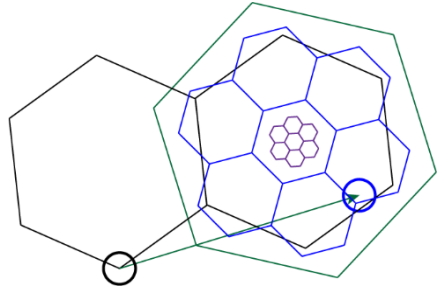

**Fig. S9| K-valley replica bands induced by Gr/WSe$_2$ superlattices.** BZs of graphite (green), bottom layer (black) WSe$_2$, mini-BZs of 5.1° tbWSe$_2$ (purple) and 11° Gr/WSe$_2$ superlattices (blue), respectively. The green arrow stands for the reciprocal vector of graphite. The black and blue circles stand for the original and replica K-valley bands, respectively.

After considering all possible situations, we can safely conclude the observed Γ-valley replica bands are intrinsic moiré bands of tbWSe$_2$.

## 6. Additional STM data

Figure S10 presents the CH-STS measured in a large bias voltage range, which covers both the conduction band and the valence band of tbWSe$_2$. The first two prominent peaks (G$_1$ & G$_2$) in the valence band can be assigned to two Γ-valley main bands (Γ$_1$ & Γ$_2$), which will be explained later. All STS data presented in the main text is obtained around G$_1$.

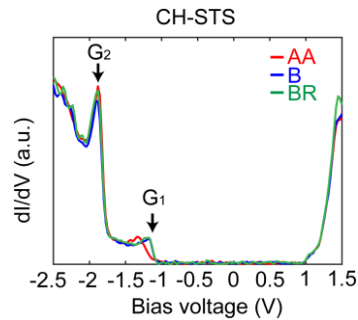

**Fig. S10| CH-STS in large bias voltage range.** CH-STS obtained from AA (red), B (blue) and BR (green) sites. G$_1$ and G$_2$ mark the peak positions in CH-STS, the signals of which are contributed from two Γ-valley main bands Γ$_1$ and Γ$_2$, respectively.

Figure S11 presents the evolution of the LDOS maps from -1.095 V (Γ-valley edge) to -1.290 V with a step size of 5 mV. The bias voltage range is highlighted by gray background in the CH-STS. The charge distribution first evolves (~30 mV below Γ-valley edge) into the kagome pattern and keeps for ~10 mV. Then it quickly changes (within 10 mV) into the honeycomb pattern and keeps for ~ 20 mV. After ~10 mV transition, the second kagome pattern appears and keeps for ~15 mV. Then the charge distribution experiences a complex evolution and finally evolves into another honeycomb pattern.

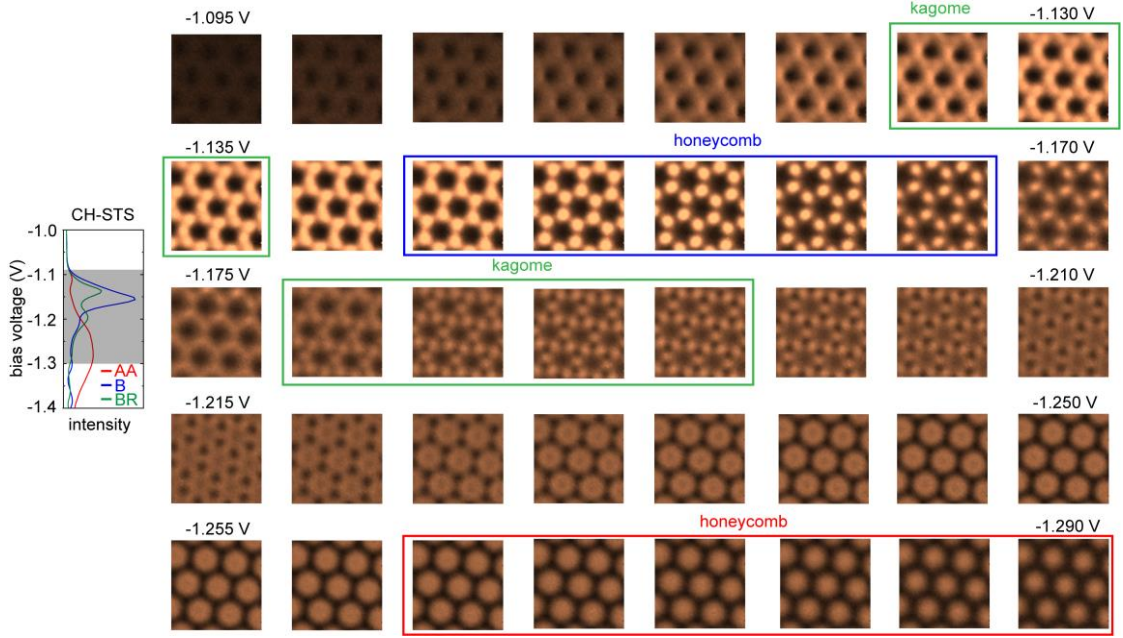

**Fig. S11| Evolution of LDOS maps.** Left: CH-STS measured at AA (red), B (blue) and BR (green) sites. The gray area shows the bias voltage range shown in the right panel. Right: from top-left to bottom-right, the evolution of LDOS maps as $V_{bias}$ changes from -1.095 V to -1.290 V with a step size of 5 mV. LDOS maps around the peaks in CH-STS are highlighted by the squares of corresponding colors.

Figure S12 presents measurements on the effective tunneling decay constants $\kappa = \sqrt{\frac{2m\phi}{\hbar^2} + k_\parallel^2}$ on different sites, where $m$ is the effective mass of tunneling electron, $\phi$ is the effective work function for the STM junction, $\hbar$ is the reduced Planck constant, and $k_\parallel$ is the in-plane momentum of tunneling electron. $\kappa$ is commonly used to discuss the origin of d$I$/d$V$ peaks in the BZ[4,5], as it can tell the relative magnitude of $k_\parallel$. To get the decay constant, the $I$-$Z$ spectroscopy was measured at a series of sample bias. For each bias voltage the set-point current was 10 pA, and $Z$ was usually swept for 3 Å. The decay constant can be deduced from the slope of ln$I$ vs $Z$ curve. Our results (Fig. S12 b) present obvious dips around -1.12 V on all high-symmetry sites, which correspond to the peak positions in constant-current STS (Fig. S12 a). Such dips indicate d$I$/d$V$ peaks around -1.12 V are contributed from tunneling electrons with smaller in-plane momentum, which is believed as an indication for the Γ-valley origin according to previous STM study on TMDs[4-6].

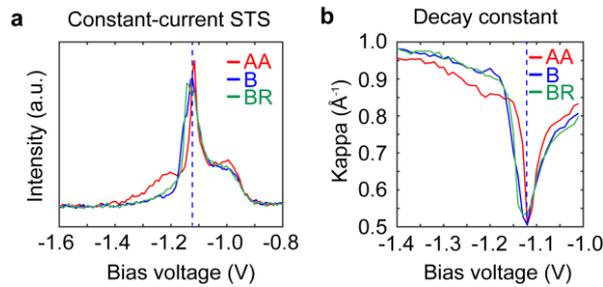

**Fig. S12| Decay constant measurement. a**, CH-STS measured on AA (red), B (blue) and BR (green)

sites. The blue dashed line indicate the peak position of B site spectrum. **b**, Decay constant as a function of bias voltage obtained from AA (red), B (blue) and BR (green) sites. The blue dashed line indicate the dip position of B site spectrum.

Figure S13 presents the STM results in another measurement, which shows consistent data of CH-STS and LDOS maps, confirming the reproducibility of our results.

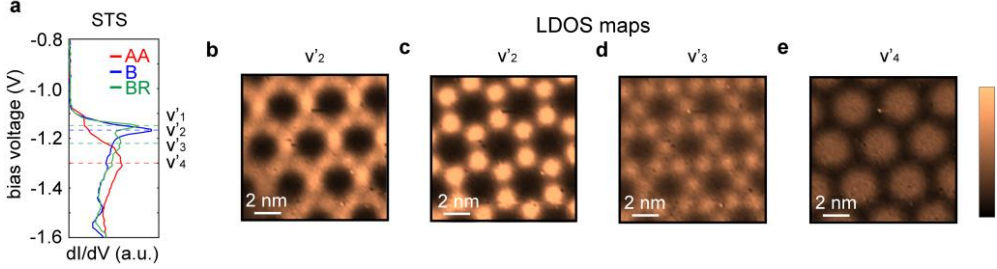

**Fig. S13| STM results in another measurement. a**, CH-STS on AA (red), B (blue) and BR (green) sites. The peak positions are indicated by dashed lines of corresponding colors, marked as v'$_1$-v'$_4$. **b-e**, LODS maps at v'$_1$-v'$_4$.

## 7. Identification of *K* and **Γ** peaks in STS measurement

Figure S14 presents the comparison between ARPES and STM results, which serve as a direct reference to distinguish the origin of peaks in STS. The good consistence between the energy distribution curve (EDC) at Γ point and the CH-STS strongly supports that the G$_1$ area in STS originates from Γ-valley top (Fig. S14 (a)). From the band dispersion along Γ-*K* direction, we obtain the energy difference between Γ and *K* valley tops is 140 meV. Thus, we can assign the small peaks in CC-STS around $V_{bias}$ = -1.0 V, which is δ*V* = 130 mV higher than G$_1$, to the *K*-valley.

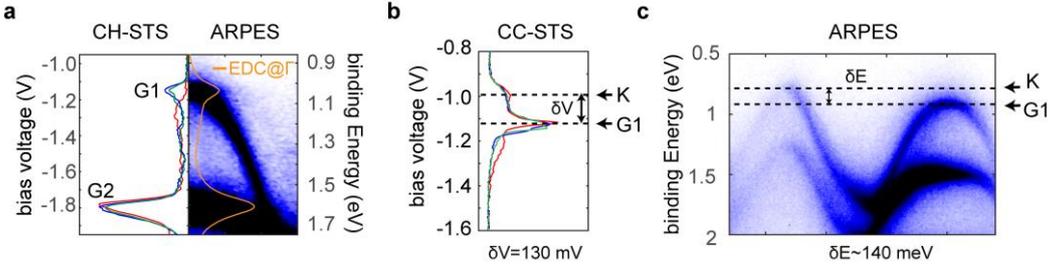

**Fig. S14| Comparison between ARPES and STM results. a**, Left: CH-STS showing the peak positions G$_1$ and G$_2$. Right: Band dispersion across Γ point. The orange spectrum shows EDC at Γ. **b**, CC-STS. δ*V* (~130 mV) shows the $V_{bias}$ bias voltage difference between *K*-valley top (marked as *K*) and Γ-valley top (marked as G$_1$). **c**, ARPES band dispersion along Γ-*K* direction. δ*E* (~140 meV) shows the binding energy difference between *K*-valley top and Γ-valley top.

In Fig. S15, we attempt to compare momentum-integrated EDCs with STS. As the integration range increases (Fig. S15 b), both peaks at G1 and G2 are broadened and down-shifted, which can be attributed to the shape of hole-pocket. From the direct comparison between STS and EDC (Fig. S15 c), we find the EDC at Zone 1 has a similar energy range as STS peaks at B and Br site, which indicates the peak in EDC include information of STS peaks.

However, a more detailed correspondence between STM and ARPES results is challenging to establish, as the quantitative interpretation of STS and EDC results is obstructed by several issues: 1. STS presents a momentum-integrated signal, while its momentum integration range is strongly dependent on the experimental conditions, as confirmed by the distinct results in CH and CC-STS; 2. ARPES intensity is dependent on the matrix-element, which is related with experimental conditions, such as photon energy, polarization, sample orientation, orbital components; 3. The energy resolution of micro-ARPES is not as good as STM, some subtle features in STS cannot be resolved in micro-ARPES.

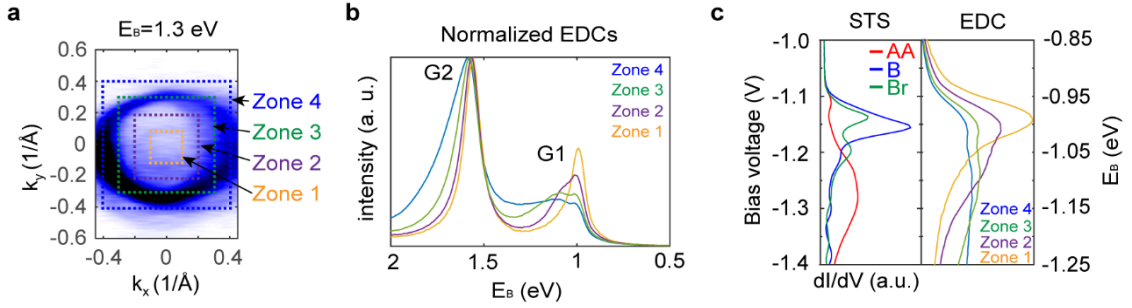

**Fig. S15| Momentum-integrated EDCs. a**, Equal energy contour to show the momentum range for integrated EDCs. Zones 1-4 present a square with side lengths of 0.2, 0.4, 0.6, 0.8 Å$^{-1}$, respectively. **a**, EDCs corresponding to Zones 1-4. The intensity has been normalized to the peak intensity of G2. **c**, Comparison between Fig. 3b and momentum-integrated EDCs.

## 8. Calculation methods

For bilayer WSe$_2$, the interlayer hopping Hamiltonian includes the interaction between the Se atoms at the interface of the bilayer, i.e., *p-p* interlayer coupling. The bonding integrals $V_{pp\sigma}$ and $V_{pp\pi}$ are taken from ref. [7], the expression of which is

$$V_{pp,b}(r) = \nu_b \exp[-(r/R_b)^{\eta_b}], \qquad (1)$$

where $b=\sigma, \pi$, and the constant values $\nu_\sigma, R_\sigma, \eta_\sigma, \nu_\pi, R_\pi, \eta_\pi$ are 2.559 eV, 3.337 Å, 4.114, -1.006 eV, 2.927 Å, 5.185, respectively. To improve the calculation accuracy of the highest valence bands at Γ, the interlayer coupling between $d_{z^2}$ and $p_z$ are included in the interlayer hopping Hamiltonian. Similar to the process in ref.[7], we obtain the empirical expressions for the bonding integrals $V_{pd\sigma}$ and $V_{pd\pi}$

$$V_{pd\sigma}(r) = A + Br, \qquad (2)$$
$$V_{pd\pi}(r) = \nu \exp[-(r/R_0)^\eta], \qquad (3)$$

where $A, B, \nu, R_0, \eta$ are 0.376 eV, -0.051 eV/Å, -0.274 eV, 4.247 Å, 5.025, respectively. In our calculations, the cut-off distance of the interlayer hopping terms is 8 Å. Figures S16 c and e show the band structures of intrinsic and twisted bilayer WSe$_2$ calculated by this TB model, respectively, which is in good agreement with the DFT results (Figs. S16 b and d).

| $E_{p_z}$ | $E_{p_x}, E_{p_y}$ | $E_{d_{z^2}}$ | $E_{d_{xz}}, E_{d_{yz}}$ | $E_{d_{xy}}, E_{d_{x^2-y^2}}$ | $V_{pp\sigma}^{intra}$ | $V_{pp\pi}^{intra}$ | $V_{pp\sigma}^{inter}$ | $V_{pp\pi}^{inter}$ |
|---|---|---|---|---|---|---|---|---|
| -4.328 | -3.778 | -2.066 | -1.075 | -2.153 | 0.713 | -0.214 | 1.344 | -0.294 |
| $V_{pd\sigma}^{1NN}$ | $V_{pd\pi}^{1NN}$ | $V_{pd\sigma}^{2NN}$ | $V_{pd\pi}^{2NN}$ | $V_{dd\sigma}$ | $V_{dd\pi}$ | $V_{dd\delta}$ | | |
| -2.258 | 1.263 | -0.505 | -0.050 | -0.869 | 1.311 | -0.045 | | |

**Table. S1| The tight-binding parameters for monolayer WSe2.** All terms are in units of eV.

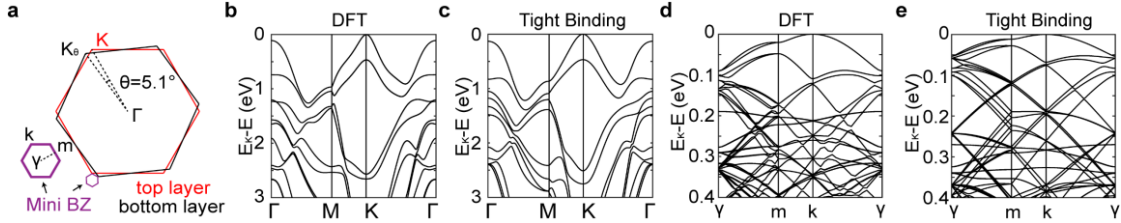

**Fig. S16| Calculated band structure. a**, Definitions of high-symmetry points used in calculation. **b-c**, The band structure of intrinsic bilayer WSe$_2$ calculated by DFT and tight binding, respectively. **d-e**, The band structure of twisted bilayer WSe$_2$ ($\theta$=5.1°) calculated by DFT and tight binding, respectively. The effect of SOC is included in all calculations.

Unfolding procedure based on the atomistic tight-binding model is widely used to simulate weighted moiré bands in the original BZ. In principle, this method can be applied to any moiré electronic systems as long as the structural relationship between the moiré superlattice and original lattice is known. The unfolding results can be presented in a 2D reciprocal space. As shown in Fig. S17 a, an energy band plotted along a k-path in the original BZ can always be folded into a path in the mini-BZ with multiple folded bands. For example, a band plotted along the ΓM direction in original BZ can be replotted as multiple folded bands along the γkmkγ direction in mini-BZ. Since the system has moiré potential, the folded bands would that are differ by integer multiples of reciprocal vectors can hybridize with each other, generating moiré bands that are different from the original folded bands. The unfolding procedure is basically to plot all the moiré bands along the k path in the original BZ with a weight assigned to each of them, and the weight is calculated by projecting these moiré bands onto the original band at each k point without the moiré potential effects. In the limit that the moiré potential goes to 0, the unfolded moiré bands is basically exactly equivalent to the original band. Such weighted moiré bands plotted along the k path in the original BZ (as shown in Fig.S17 b) can illustrate the moiré potential effects in a clearer way.

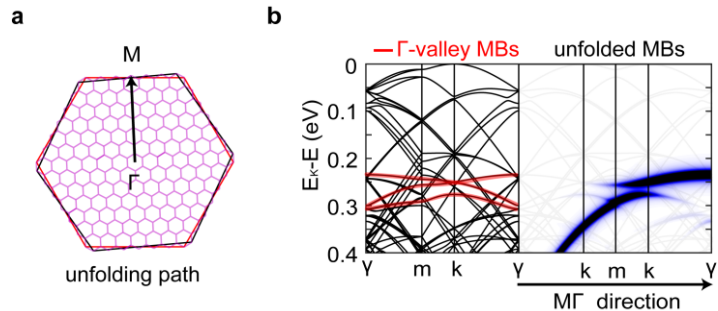

**Fig. S17| Unfolding process. a**, The equivalent path in BZ and mini-BZ. **b** Unfolding process along the equivalent path.

## 9. Calculated charge distribution

Figure S18 presents the calculated charge distribution from Γ-valley edge to the third Γ-valley moiré band. Here, e1, e3 and e4 are the binding energies of the top-three Γ-valley bands at the $\gamma$ point. e2 is the binding energy of the crossing point of the top-two Γ-valley bands. The evolution of calculated charge distribution is qualitatively consistent with that of LDOS maps (Fig. S11, Fig. S18 c).

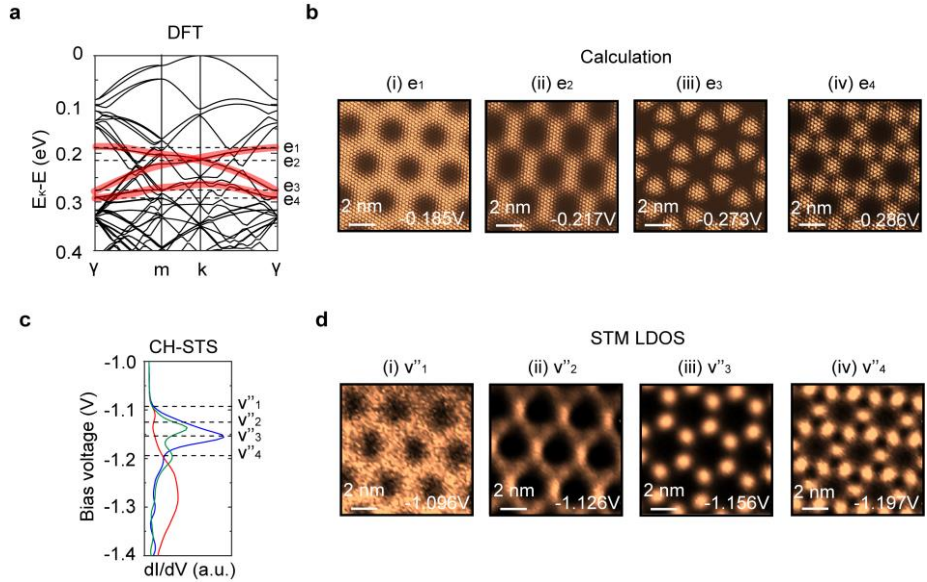

**Fig. S18| Calculation vs STM. a**, The band structure of twisted bilayer $WSe_2$ ($\theta$=5.1°) calculated by DFT. The top-most three Γ-valley moiré bands are highlighted in red **b**, Calculated charge distributions corresponding to (i)-(v) e1-e4. **c**, CH-STS. **d**, LODS maps at $V_{bias}$ (i)-(v) v''1-v''4. The intensity in the calculated charge distributions and the LDOS maps has been normalized for the better comparison.

## 10. Estimated gap size at the replicated crossing point

As shown in Fig. S19 a, according to our calculation, the gap size for the rigidly twisted tbWSe$_2$ model is merely ~2 meV, and even in a fully relaxed lattice (Fig. S19 b), it is ~12 meV. Thus, in reality, the gap is likely to be somewhere between 2~12 meV, which is unfortunately smaller than our energy resolution, thus may not be decerned in our measurement (e.g. in Fig. 2e).

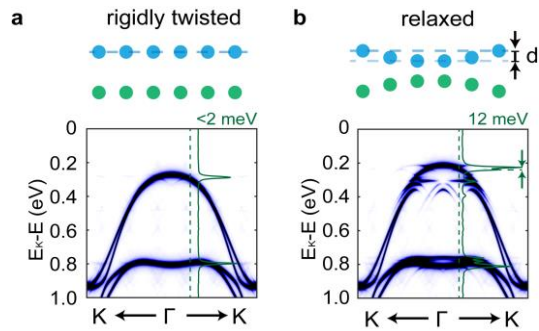

**Fig. S19| Simulation of gap size. a**, Estimated gap size of rigidly twisted (without relaxation) tbWSe$_2$ superlattice. **b**, Estimated gap size of relaxed tbWSe$_2$ superlattice. The green curves are energy distribution curves at the replicated crossing point (indicated by green dashed lines).

## 11. Situation when Γ-valley is higher

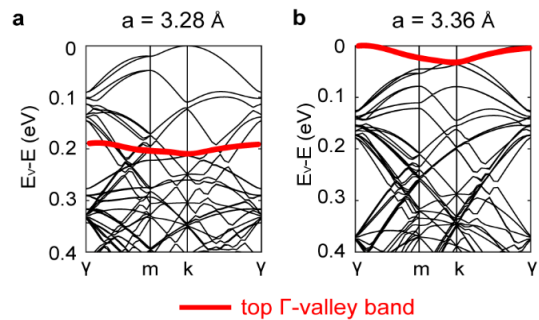

**Fig. S20| DFT calculated moiré bands with different WSe$_2$ lattice constants. a**, a = 3.28 Å. **b**, a = 3.36 Å.

Here, we show DFT-calculated moiré bands with different WSe$_2$ lattice constant (Fig. S20). In the calculation with a = 3.28 Å, the top-most Γ-valley moiré band is located ~ 0.2 eV below the K-valley top. When the lattice constant increases to 3.36 Å while keeping the volume unchanged, the top-most Γ-valley bands are pushed up to the valence band edge. These discoveries indicate, when tbWSe$_2$ is under sufficient pressure (or in-plane strain, or lattice reconstruction), its electrical properties in p-doped region could be dominated by the Γ-valley.